\documentclass[conference]{IEEEtran}
\usepackage{cite}

\usepackage{graphicx}

\usepackage{bm}
\usepackage{amsfonts}

\usepackage{amssymb}

\usepackage{color}

\usepackage[outdir=./]{epstopdf}
\ifCLASSINFOpdf
\else
\fi
\usepackage{amsmath}
\usepackage{algorithm}
\usepackage{algorithmic}

\usepackage{booktabs}

\ifCLASSOPTIONcompsoc
 \usepackage[caption=false,font=normalsize,labelfont=sf,textfont=sf]{subfig}
\else
 \usepackage[caption=false,font=footnotesize]{subfig}
\fi
\usepackage{stfloats}
\fnbelowfloat
\begin{document}
%
\title{Radar-assisted Predictive Beamforming for Vehicle-to-Infrastructure Links}

\author{\IEEEauthorblockN{Fan Liu\IEEEauthorrefmark{1},
Weijie Yuan\IEEEauthorrefmark{2},
Christos Masouros\IEEEauthorrefmark{1}
and Jinhong Yuan\IEEEauthorrefmark{2}
\IEEEauthorblockA{\IEEEauthorrefmark{1}Department of Electronic and Electrical Engineering, University College London, London, UK}
\IEEEauthorblockA{\IEEEauthorrefmark{2}School of Electrical Engineering and Telecommunications, University of New South Wales, Sydney, Australia}
Email: $\left\{\text{fan.liu, c.masouros} \right\}$@ucl.ac.uk, $\left\{\text{weijie.yuan, j.yuan} \right\}$@unsw.edu.au}
}


%


\maketitle

\begin{abstract}
In this paper, we propose a radar-assisted predictive beamforming design for vehicle-to-infrastructure (V2I) communication by relying on the joint sensing and communication functionalities at road side units (RSUs). We present a novel extended Kalman filtering (EKF) framework to track and predict kinematic parameters of the vehicle. By exploiting the radar functionality of the RSU we show that the communication beam tracking overheads can be drastically reduced. Numerical results have demonstrated that the proposed radar-assisted approach significantly outperforms the communication-only feedback based technique in both the angle tracking and the downlink communication.
\end{abstract}


%
\IEEEpeerreviewmaketitle

\section{Introduction}
Sensing and communication functionalities will be intertwined with each other in the future vehicle-to-everything (V2X) network. To provide both high-accuracy localization and high-throughput communication services for autonomous vehicles, 5G Millimeter Wave (mmWave) and massive multi-input-multi-output (mMIMO) technologies have been proposed as promising solutions. In contrast to their 4G counterparts which have only basic localization capability, the large bandwidth available at the mmWave spectrum together with the mMIMO antenna array offer a range resolution at the order of 10cm, and an angular resolution that is less than $1^\circ$ \cite{8246850}. Furthermore, the 5G techniques will also allow Gbps data transmission at a latency less than 1s \cite{8246850}, which are beneficial for high-mobility vehicular applications.
\\\indent In light of the above background, the need for joint sensing and communication designs naturally emerges in the vehicular network. It has been shown that by employing a single device for the dual purposes of sensing and communication, the computational and hardware costs can be considerably reduced \cite{Joint_TCOM}. Moreover, the overall system performance can be improved via the cooperation between the two functionalities \cite{Joint_TCOM}. For these reasons, the research of dual-functional radar-communication (DFRC) systems has recently attracted substantial attentions from both academia and industry.
\\\indent Aiming for combining radar and communication signals on the temporal and the spectral domains, early works on DFRC have explored the possibility of modulating frequently-used radar waveforms, e.g., chirp signal or spread-spectrum sequences, with communication data symbols. To exploit the favorable time-frequency decoupling property of the Orthogonal Frequency Division Multiplexing (OFDM) waveforms, the pioneering paper \cite{sturm2011waveform} proposed to employ OFDM communication signals for radar detection, where the delay and the Doppler parameters can be estimated independently. More relevant to this work, the spatial processing aspect for DFRC systems has been extensively investigated thanks to the development of the multi-antenna technology. By resorting to the high degrees-of-freedom (DoFs) of the MIMO systems, a straightforward DFRC approach is to employ the main-beam of the MIMO radar for target detection, while using the sidelobes for conveying useful information to the communication users in a line-of-sight (LoS) channel \cite{7347464}. To further enhance the system performance, the recent treatises \cite{8288677,8386661} have proposed novel DFRC beamforming and waveform design techniques by providing communication service in the non-line-of-sight (NLoS) channels.
\\\indent While the aforementioned schemes have implemented DFRC functionalities on lower frequency bands, e.g., sub-6 GHz, they are difficult to be extended to the V2X applications that operate in the mmWave band, where specific mmWave channel models and vehicular constraints should be taken into account. In \cite{7888145}, a radar-aided beam alignment method has been developed for mmWave vehicle-to-infrastructure (V2I) communications, where an extra radar device has been deployed on the road infrastructure in addition to the communication system, which inevitably leads to high hardware costs compared to the DFRC designs. In view of this, a mmWave DFRC system has been proposed in \cite{8642926} for integrating the bi-static automotive radar and vehicle-to-vehicle (V2V) communications, which, however, does not address the issue of beam tracking under vehicular scenarios with high mobility.
\\\indent Building upon the communication-only protocols, conventional mmWave beam tracking approaches require the transmitter to send pilots to the receiver; the receiver then estimates the angle and feeds it back to the transmitter \cite{8809900}. It is worth pointing out that for high-mobility scenarios, beam tracking is not sufficient in general. More importantly, the transmitter should be able to \emph{predict} the beam, given the critical latency requirement. By realizing this, the state-of-the-art techniques have employed Kalman filtering for beam prediction and tracking based on the feedback protocol mentioned above \cite{7905941}. Typically, these approaches utilize only a small number of pilots for beam tracking, resulting in limited matched filtering gain for angle estimation. Moreover, inserting pilot symbols into the communication block may cause significant overhead, and thus reduces the transmission rate of the useful information.
\\\indent To cope with the above issues, we propose in this paper a novel predictive beamforming design for the V2I communication link by the DFRC signaling, where the reflected echo signal is exploited for both beam tracking and prediction instead of using the conventional uplink feedback method. Since the whole downlink communication block is leveraged for accomplishing the dual tasks of radar target detection and communication data transmission, no extra downlink pilots are required. Moreover, the operation of matched-filtering/pulse compression would bring considerable gain in the receive signal-to-noise ratio (SNR). In line with the spirit of joint sensing and communication, we further develop an extended Kalman filtering (EKF) scheme for tracking and predicting the motion parameters of the vehicle. Simulation results show that the proposed method is significantly superior to the conventional feedback based technique in both localization and communication performances.
\section{General Framework}
We consider a mmWave mMIMO RSU with a uniform linear array (ULA), which serves a single vehicle on the road as depicted in Fig. 1. To communicate with the RSU, the vehicle is also equipped with an MIMO array at both sides of the body. For notational simplicity, and without loss of generality, we assume that the vehicle is driving along a straight road that is parallel to the antenna array of the RSU, and that the RSU communicates with the vehicle via a LoS channel. The discussion of NLoS channels is designated to our future work. In what follows, we will firstly introduce the general framework, and then the detailed signal model.
\\\indent \emph{Remark 1:} Note that the ULA of the RSU can be adjusted to be paralleled to the road, where small mismatches are allowed. In fact, alternative relative directions can be straightforwardly accommodated by adding a fixed offset to the tracked angles. We note here that this offset can be easily calibrated since it is fixed and is known to the RSU. As a result, our proposed techniques can be applied without any changes.
\\\indent Let us denote the angle, the distance and the velocity of the vehicle relative to the RSU's array as $\theta\left(t\right)$, $d\left(t\right)$, $v\left(t\right)$, respectively. Further, the angle of the RSU relative to the vehicle is denoted as $\phi\left(t\right)$. Note that all the parameters are functions of time $t \in \left[0,T\right]$, with $T$ being the maximum time duration of interest. It then follows that $\phi\left(t\right) = \theta\left(t\right)$, given the parallel driving directions of the vehicles relative to the RSU's antenna array. We therefore omit $\phi$ in the remainder of the paper. For notational convenience, we discretize the time period $T$ into several small time-slots with a length of $\Delta T$, and denote $\theta_{n}$, $d_{n}$ and $v_{n}$ as the motion parameters at the \emph{n}th epoch for each vehicle. Following the standard assumption in the literature \cite{7905941}, we assume that the motion parameters keep constant within $\Delta T$.
\\\indent \emph{1) Initial Estimation}
\\\indent Our proposed scheme is initialized by letting the RSU estimate the parameters of the vehicle that enters into the coverage of interest. In this stage, the RSU can either act as a pure mono-static radar, which infers the initial vehicle parameters $\theta_{0}$, $d_{0}$ and $v_{0}$ from the reflected echoes, or to obtain these estimates simply via conventional uplink training. Here we note that while the RSU is only able to attain the radial velocity $v_{n}^{R}$ by estimating the Doppler frequency, it can infer the overall velocity as $v_{n} = v_{n}^{R}/\cos\theta_{n}$.
\\\indent\emph{2) State Prediction}
\\\indent With the estimates of the motion parameters $\hat \theta_{n-1}$, $\hat{d}_{n-1}$ and ${\hat v}_{n-1}$ at the $\left(n-1\right)$th epoch, the RSU performs one- and two-step predictions of the angle parameters, respectively. For the purpose of sensing, the RSU will also need to perform one-step prediction for other motion parameters, i.e., distance and velocity. At the $n$th epoch, the RSU formulates the transmit beam towards the vehicle by using the one-step predictions ${\hat \theta _{{n\left| {n - 1} \right.}}}$. Within this beam, the RSU will send a joint radar-communication signal that contains the information of the two-step predictions ${{\hat \theta }_{n + 1\left| {n - 1} \right.}}$. Once the vehicle receives the information, it will correspondingly formulate the receive beam at the $\left(n+1\right)$th epoch based on the predicted angle. The reason for using the two-step prediction at the vehicle is that the one-step predicted angle ${\hat \theta _{{n\left| {n - 1} \right.}}}$ would be outdated at the $\left(n+1\right)$th epoch. Note that the predictions are performed by using the kinematic equations of the vehicle. The transmit beam of the RSU and the receive beam of the vehicle will be aligned with each other if the estimation and prediction are sufficiently accurate.
\\\indent\emph{3) Vehicle Tracking}
\\\indent At the $n$th epoch, the signal transmitted by the RSU is partially reflected by the body of the vehicle, and is also partially received by the vehicle's antenna array. As discussed above, for each vehicle, the data sequence received contains the predicted angular information for the $\left(n+1\right)$th epoch, which will be exploited for receive beamforming at the vehicles. On the other hand, the RSU receives the echoes reflected by the vehicle, and estimates $\theta_{n},v_{n}$ and $d_{n}$, which are used to refine the predicted parameters at the $n$th epoch. The refined state parameters are then used as the inputs of the predictor for the $\left(n+1\right)$th and $\left(n+2\right)$th epoches at the RSU.
\\\indent For clarity, we summarize the above procedure in Fig. 1. It can be observed that by iteratively performing beam prediction and beam tracking, the RSU is able to serve multiple vehicles simultaneously. Moreoever, with the aid of the radar functionality built in the RSU, one can avoid frequent feedbacks between the RSU and vehicle. This is evidently shown in Fig. 1 that the uplink feedback from the vehicles to the RSU are replaced by the echo signal. In this sense, the beam information can be extracted by the echo signal, and all the uplink resources can be used to transfer useful data rather than the feedback information.
\begin{figure}[!t]
    \centering
    \includegraphics[width=0.8\columnwidth]{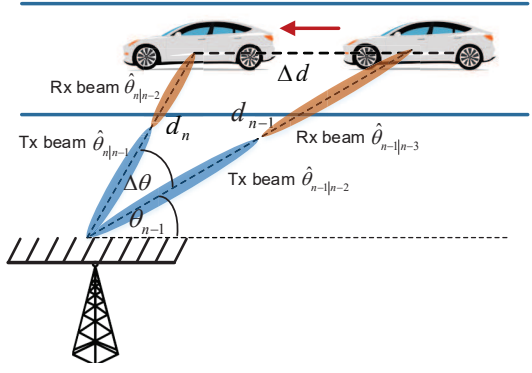}
    \caption{V2I state evolution model.}
    \label{fig:1}
\end{figure}
\section{System Model}
Based on the above discussion, it is clear that the dominating complexity in the signal processing is at the RSU's side. Moreover, the initial estimation can be simply done by conventional radar signal processing or uplink beam training. Given the aforementioned reasons, we will focus on the prediction and tracking stages at the RSU's side. In this section, we develop the measurement model at the RSU by using radar signal processing techniques.
\subsection{Radar Signal Model}
Let us denote the downlink DFRC stream transmitted at the $n$th epoch and time $t$ as ${{s}_n}\left( t \right)$. The transmitted signal can be expressed as
\begin{equation}\label{eq1}
  {{\mathbf{\tilde s}}}_n\left(t\right) = {\mathbf{f}}_n{{s}}_n\left(t\right) \in \mathbb{C}^{N_t \times 1},
\end{equation}
where ${\mathbf{f}}_n \in \mathbb{C}^{N_t \times 1}$ is the transmit beamforming vector, with $N_t$ being the number of transmit antennas. Accordingly, the reflected echo signal received at the RSU can be given in the form
\begin{equation}\label{eq7}
{{\mathbf{r}}_{n}}\left( t \right) =
\kappa\sqrt {p_{n}}{\beta _{n}}{e^{j{2\pi \mu_{n}}t}}{\mathbf{b}}\left( {{\theta _{n}}} \right){{\mathbf{a}}^H}\left( {{\theta _{n}}} \right){{\mathbf{f}}_{n}}{{s}}_{n}\left( {t - {\tau _{n}}} \right)
+ \mathbf{z}_{n}\left( t \right).
\end{equation}
where $p_{n}$ is the transmit power at the $n$th epoch, $\kappa = \sqrt{N_tN_r}$ is the array gain factor, with $N_r$ being the number of receive antennas, $\mathbf{z}_n\left( t \right) \in \mathbb{C}^{N_r \times 1}$ represents the complex additive white Gaussian noise with zero mean and variance of $\sigma^2$, $\beta_{n}$, $\mu_{n}$ and $\tau_{n}$ denote the reflection coefficient, the Doppler frequency and the time-delay for the vehicle. Given the distance $d_{n}$, the reflection coefficient can be expressed as
\begin{equation}\label{eq3}
  {\beta _{n}} = {\varepsilon _{n}}{\left( {2{d_{n}}} \right)^{ - 1}},
\end{equation}
where $\varepsilon _{n}$ is the complex radar cross-section (RCS) of the vehicle at the $n$th epoch. We assume that the RCS of the vehicle keeps constant during the period $T$, i.e., ${\varepsilon _{n}} = {\varepsilon _{n-1}} = ...= {\varepsilon _{0}}, \forall n$, which corresponds to a Swerling I target \cite{richards2005fundamentals}. In (\ref{eq7}), ${\mathbf{a}}\left( {{\theta}} \right)$ and ${\mathbf{b}}\left( {{\theta}} \right)$ are transmit and receive steering vectors of the antenna array of the RSU, which are expressed as
\begin{equation}\label{eq4}
{\mathbf{a}}\left( \theta  \right) = \sqrt{\frac{1}{N_t}}{\left[ {1,{e^{-j\pi \cos \theta }},...,{e^{-j\pi \left( {{N_t} - 1} \right)\cos \theta }}} \right]^T},
\end{equation}
\begin{equation}\label{eq5}
{\mathbf{b}}\left( \theta  \right) = \sqrt{\frac{1}{N_r}}{\left[ {1,{e^{-j\pi \cos \theta }},...,{e^{-j\pi \left( {{N_r} - 1} \right)\cos \theta }}} \right]^T},
\end{equation}
where we assume half-wavelength antenna spacing for the ULA.
\\\indent The beamforming vector $\mathbf{f}_n$ is designed based on the prediction of the angle, which is
\begin{equation}\label{eq6}
  {{\mathbf{f}}_{n}} = \mathbf{a}\left( {{{\hat \theta }_{{n\left| {n - 1} \right.}}}} \right),
\end{equation}
where ${{{\hat \theta }_{{n\left| {n - 1} \right.}}}} $ is the one-step predicted angle for vehicle at the $n$th epoch. By employing the design in (\ref{eq6}), the RSU will formulate a transmit beam towards the predicted direction to track the vehicle.
\\\indent In (\ref{eq7}), the transmit signal-to-noise ratio (SNR) is defined as $ \frac{{{p_{n}}}}{{{\sigma ^2}}}$. By matched-filtering (\ref{eq7}) with a delayed and Doppler-shifted version of ${{s}}_{n}\left( {t} \right)$, one can estimate the delay $\tau_{n}$ and the Doppler frequency $\mu_{n}$. Compensating (\ref{eq7}) using these estimates yields the measurement model for the angle $\theta_{n}$ and the reflection coefficient $\beta_{n}$ as
\begin{equation}\label{eq8}
\begin{gathered}
  {{\mathbf{\tilde r}}_{n}} = \kappa{\beta _{n}}{\mathbf{b}}\left( {{\theta _{n}}} \right){{\mathbf{a}}^H}\left( {{\theta _{n}}} \right){{\mathbf{f}}_{n}} + {{\mathbf{z}}_{\theta}} \hfill \\
   = \kappa{\beta _{n}}{\mathbf{b}}\left( {{\theta _{n}}} \right){{\mathbf{a}}^H}\left( {{\theta _{n}}} \right){\mathbf{a}}\left( {{{\hat \theta }_{n\left| {n - 1} \right.}}} \right) + {{\mathbf{z}}_{\theta}}, \hfill \\
\end{gathered}
\end{equation}
where ${\mathbf{z}}_{\theta}$ denotes the measurement noise normalized by the transmit power $p_{n}$ and the matched-filtering gain $G$, with zero mean and variance of $\sigma_1^2$. Note here that $G$ is the SNR gain brought by the matched-filtering operation, which typically equals to the energy of $s_{n}\left(t\right)$. Furthermore, the measurement models of the distance $d_{n}$ and the velocity $v_{n}$ are given as
\begin{equation}\label{eq9}
  {\tau _n} = {{2{d_n}} \mathord{\left/
 {\vphantom {{2{d_n}} c}} \right.
 \kern-\nulldelimiterspace} c} + {z_\tau },
\end{equation}
\begin{equation}\label{eq10}
  {\mu _n} = {{2{v_n}\cos {\theta _n}{f_c}} \mathord{\left/
 {\vphantom {{2{v_n}\cos {\theta _n}{f_c}} c}} \right.
 \kern-\nulldelimiterspace} c} + {z_f},
\end{equation}
where $f_c$ and $c$ represent the carrier frequency and the speed of light, respectively, ${z_\tau }$ and $z_f$ denote the measurement Gaussian noise with zero mean and variance of $\sigma_2^2$ and $\sigma_3^2$, respectively. Note that the round-trip is twice the distance from the RSU to the vehicle, and the Doppler frequency relies on the radial velocity ${v_{n}}\cos {\theta _{n}}$. Moreover, we remark here that the variances of the measurement noises are inversely proportional to the receive signal-to-noise ratio (SNR) of (\ref{eq7}) \cite{kay1998fundamentals}, i.e.,
\begin{equation}
  \sigma _1^2 = \frac{{{a_1^2}{\sigma ^2}}}{{G{p_{n}}}},\sigma _i^2 = \frac{{{a_i^2}{\sigma ^2}}}{{G{\kappa ^2}{{\left| {{\beta _{n}}} \right|}^2}{\left|\delta _{n}\right|}^2{p_{n}}}},i = 2,3,
\end{equation}
where ${\delta _{n}} = {{\mathbf{a}}^H}\left( {{\theta _{n}}} \right){\mathbf{a}}\left( {{{\hat \theta }_{n\left| {n - 1} \right.}}} \right)$ represents the beamforming gain factor, whose modulus equals to 1 if the predicted angle perfectly matches the real angle, and is less than 1 otherwise. Note that $\sigma_2^2$ and $\sigma_3^2$ are determined by the transmit power $p_{n}$, the matched filtering gain $G$, the array gain $\kappa$, the beamforming gain ${\delta _{n}}$ as well as the strength of the reflected signal. Nevertheless, $\sigma _1^2$ is only determined by the transmit power $p_{n}$ and the matched filtering gain $G$, since $\kappa$, $\beta_{n}$ and ${\delta _{n}}$ are already contained in (\ref{eq8}). Finally, $a_i, i= 1,2,3$ are constants related to the system configuration, signal designs as well as the specific signal processing algorithms.
\subsection{Communication Model}
As shown in Fig. 1, at the $n$th epoch, the vehicle receives the signal from the RSU by using a receive beamformer $\mathbf{w}_{n}$, yielding
\begin{equation}\label{eq11}
  {c_{n}}\left( t \right) =
    {\tilde\kappa}\sqrt{p_{n}}{\alpha _{n}}{\mathbf{w}}_{n}^H{\mathbf{u}}\left( {{\theta _{n}}} \right){{\mathbf{a}}^H}\left( {{\theta _{n}}} \right){{\mathbf{f}}_{n}}{s_{n}}\left( t \right) + {z_c}\left( t \right),
\end{equation}
where ${{z}_c}\left( t \right)$ is the zero-mean Gaussian noise with variance ${\sigma_C^2}$, ${s}_{n}\left(t\right)$ denotes the DFRC stream transmitted from the RSU to thevehicle, ${\alpha _{n}}$ denotes the communication channel coefficient, which is different from the radar reflection coefficient ${\beta _{n }}$, ${\mathbf{u}}\left(\theta\right)$ represents the steering vector of the vehicle's antenna array, and is similarly defined as in (\ref{eq4}) and (\ref{eq5}) with $M$ antennas.
\\\indent As discussed in the above, the receive beamformer should be formulated based on the two-step prediction of the angle parameter, since the one-step predicted information would be outdated for receive beamforming at the vehicle. This is expressed as
\begin{equation}\label{eq12}
{{\mathbf{w}}_{n}} = {\mathbf{u}}\left( {{{\hat\theta} _{n\left| {n - 2} \right.}}} \right).
\end{equation}
Assume that the DFRC stream ${s}_{n}\left(t\right)$ has a unit power, then the receive SNR for the vehicle at the $n$th epoch is obtained as
\begin{equation}\label{eq13}
{\operatorname{SNR} _n} = {{{p_n}{{\left| {\tilde \kappa {\alpha _n}{\mathbf{w}}_n^H{\mathbf{u}}\left( {{\theta _n}} \right){{\mathbf{a}}^H}\left( {{\theta _n}} \right){{\mathbf{f}}_n}} \right|}^2}} \mathord{\left/
 {\vphantom {{{p_n}{{\left| {\tilde \kappa {\alpha _n}{\mathbf{w}}_n^H{\mathbf{u}}\left( {{\theta _n}} \right){{\mathbf{a}}^H}\left( {{\theta _n}} \right){{\mathbf{f}}_n}} \right|}^2}} {\sigma _C^2}}} \right.
 \kern-\nulldelimiterspace} {\sigma _C^2}}.
\end{equation}
Accordingly, the achievable rate at the $n$th epoch is given as
\begin{equation}\label{eq14}
  {R_n} = {\log _2}\left( {1 + {{\operatorname{SNR} }_n}} \right).
\end{equation}
Following the standard assumption in the literature, the LoS channel coefficient $\alpha_{n}$ is given as \cite{8675440}
\begin{equation}\label{eq16}
  {\alpha _{n}} = {\tilde \alpha}d_{n}^{ - 1}{e^{j\frac{{2\pi }}{\lambda }{d_{n}}}} = {{\tilde \alpha}}d_{n}^{ - 1}{e^{j\frac{{2\pi {f_c}}}{c}{d_{n}}}},
\end{equation}
where ${\tilde \alpha}d_{n}^{ - 1}$ is the path-loss of the channel, with ${\tilde \alpha}$ being the channel power gain at the reference distance $d_0 = 1\text{m}$, $\frac{{2\pi }}{\lambda }{d_{n}}$ is the phase of the LoS channel, with $\lambda = \frac{{{f_c}}}{c}$ being the wavelength of the signal.
\subsection{State Evolution Model}
Our goal is to track the variation of the angle and distance of the vehicle by processing the measured signals in (\ref{eq8}), (\ref{eq9}) and (\ref{eq10}), which are determined by the kinematic equations of the vehicle. First of all, based on the geometric relations shown in Fig. 1, we have
\begin{equation}\label{eq17}
\left\{ \begin{gathered}
  d_n^2 = d_{n - 1}^2 + \Delta {d^2} - 2{d_{n - 1}}\Delta d\cos {\theta _{n - 1}}, \hfill \\
  \frac{{\Delta d}}{{\sin \Delta \theta }} = \frac{{{d_n}}}{{\sin {\theta _{n - 1}}}}, \hfill \\
\end{gathered}  \right.
\end{equation}
where $\Delta d = {v_{n-1}}\Delta T$, $\Delta \theta = \theta_n - \theta_{n-1}$. It is challenging to analyze the evolution model directly given the high non-linear nature of (\ref{eq17}). Let us assume that the vehicle is moving at an approximately constant speed, such that $v_n \approx v_{n-1}$. Moreover, the reflection coefficient $\beta_n$ is solely dependent on the distance $d_n$ based on (\ref{eq3}). Given the fact that the position of the vehicle will not change too much within two epoches, and by using some simple algebraic manipulations, one can obtain an approximated state evolution model as follows
\begin{equation}\label{eq29}
\left\{ \begin{gathered}
  {\theta _n} = {\theta _{n - 1}} + d_{n - 1}^{ - 1}{v_{n - 1}}\Delta T\sin {\theta _{n - 1}} + {\omega _\theta }, \hfill \\
  {d_n} = {d_{n - 1}} - {v_{n - 1}}\Delta T\cos {\theta _{n - 1}} + {\omega _d}, \hfill \\
  {v_n} = {v_{n - 1}} + {\omega _v}, \hfill \\
  {\beta _n} = {\beta _{n - 1}}\left( {1 + d_{n - 1}^{ - 1}{v_{n - 1}}\Delta T\cos {\theta _{n - 1}}} \right) + {\omega _\beta }, \hfill \\
\end{gathered}  \right.
\end{equation}
where $\omega_\theta$, $\omega_d$, $\omega_v$ and ${\omega_{\beta} }$ denote the corresponding noises, which are assumed to be zero-mean Gaussian distributed with variances of ${\sigma _\theta^2,\sigma _d^2,\sigma _v^2}$ and $\sigma_{\beta}^2$, respectively. Here we omit the details of the mathematical derivation due to the strict page limit. We highlight that these noises are generated by approximation and other systematic errors, which are irrelevant to the measurement SNR defined in the Sec. III-A.
\section{Extended Kalman Filtering}
In this section, we propose a Kalman filtering scheme for beam prediction and tracking. Due to the nonlinearity in the measurement and the state evolution models, the linear Kalman filtering (LKF) can not be directly applied. We therefore consider an EKF approach that performs local linearization for nonlinear models. By denoting the state variables as ${\mathbf{x}} = {\left[ {\theta ,d,v,\beta} \right]^T}$ and the measured signal vector as ${\mathbf{y}} = {\left[ {{{{\mathbf{\tilde r}}}^T},\tau ,\mu } \right]^T}$, the models developed in (\ref{eq29}) and (\ref{eq8})-(\ref{eq10}) can be recast in compact forms as
\begin{equation}\label{eq30}
\left\{ \begin{gathered}
  {\text{State}}\;{\text{Evolution}}\;{\text{Model: }}{{\mathbf{x}}_n} = {\mathbf{g}}\left( {{{\mathbf{x}}_{n - 1}}} \right) + {{\bm{\omega }}_n}, \hfill \\
  {\text{Measurement}}\;{\text{Model:}}\;{{\mathbf{y}}_n} = {\mathbf{h}}\left( {{{\mathbf{x}}_n}} \right) + {{\mathbf{z}}_n}, \hfill \\
\end{gathered}  \right.
\end{equation}
where $\mathbf{g}\left(\cdot\right)$ is defined in (\ref{eq29}), with $\bm{\omega} = {\left[ {{\omega _\theta },{\omega _d},{\omega _v},{\omega _\beta }} \right]^T}$ being the noise vector that is independent to ${\mathbf{g}}\left( {{{\mathbf{x}}_{n - 1}}} \right)$. Similarly, $\mathbf{h}\left(\cdot\right)$ is defined as (\ref{eq8})-(\ref{eq10}), with ${\mathbf{z}} = {\left[ {{\mathbf{z}}_\theta ^T,{z_\tau },{z_f}} \right]^T}$ being the measurement noise that is independent to ${\mathbf{h}}\left( {{{\mathbf{x}}_n}} \right)$. As considered above, both $\bm{\omega}$ and $\mathbf{z}$ are zero-mean Gaussian distributed, with covariance matrices being expressed as
\begin{equation}\label{eq31}
  {{\mathbf{Q}}_s} = \operatorname{diag} \left( {\sigma _\theta^2,\sigma _d^2,\sigma _v^2,\sigma _{\beta}^2}\right),
\end{equation}
\begin{equation}\label{eq32}
  {{\mathbf{Q}}_m} = \operatorname{diag} \left( {\sigma _1^2{{\mathbf{1}}_{{N_r}}^T},\sigma _2^2,\sigma _3^2}\right),
\end{equation}
where $\mathbf{1}_{N_r}$ denotes a size-$N_r$ all one column vector. In order to linearize the models, the Jacobian matrices for both $\mathbf{g}\left(\mathbf{x}\right)$ and $\mathbf{h}\left(\mathbf{x}\right)$ need to be computed. By simple algebraic derivation, the Jacobian matrix for $\mathbf{g}\left(\mathbf{x}\right)$ can be given as
\begin{equation}\label{eq33}
\begin{gathered}
  \frac{{\partial {\mathbf{g}}}}{{\partial {\mathbf{x}}}} = \hfill \\
  \left[ {\begin{array}{*{20}{c}}
  {1 + \frac{{v\Delta T\cos \theta }}{d}}&{ - \frac{{v\Delta T\sin \theta }}{{{d^2}}}}&{\frac{{\Delta T\sin \theta }}{d}}&0\\
  \scriptstyle{v\Delta T\sin \theta }&1&\scriptstyle{ - \Delta T\cos \theta }&0 \\
  0&0&1&0 \\
    - \frac{\beta{v\Delta T\sin \theta }}{d}&-\frac{{\beta v\Delta T\cos \theta }}{{{d^2}}}&\frac{{\beta \Delta T\cos \theta }}{d}&1 + \frac{{v\Delta T\cos \theta }}{d}
\end{array}} \right].
\end{gathered}
\end{equation}
For $\mathbf{h}\left(x\right)$, let us denote
\begin{equation}\label{eq34-1}
  {\bm{\eta }}\left( {\beta ,\theta } \right) = \kappa\beta {\mathbf{b}}\left( \theta  \right){{\mathbf{a}}^H}\left( \theta  \right){\mathbf{a}}\left( {\hat \theta } \right),
\end{equation}
where $\hat \theta$ is a prediction for $\theta$. The Jacobian matrix for $\mathbf{h}\left(x\right)$ can be then given by
\begin{equation}\label{eq34}
\frac{{\partial {\mathbf{h}}}}{{\partial {\mathbf{x}}}} = \left[ {\begin{array}{*{20}{c}}
  {\frac{{\partial {\bm\eta} }}{{\partial \theta }}}&0&0&{\frac{{\partial {\bm\eta} }}{{\partial {\beta }}}} \\
  0&{\frac{2}{c}}&0&0 \\
  { - \frac{{2v\sin \theta }}{c}}&0&{\frac{{2{f_c}\cos \theta }}{c}}&0
\end{array}} \right].
\end{equation}
where ${\frac{{\partial {\bm{\eta }}}}{{\partial \theta }}}$ and ${\frac{{\partial {\bm{\eta }}}}{{\partial {\beta} }}}$ can be readily obtained by using (\ref{eq4}), (\ref{eq5}) and (\ref{eq34-1}).
\\\indent We are now ready to present the EKF technique. Following the standard procedure of Kalman filtering \cite{kay1998fundamentals}, the state prediction and tracking design is summarized as follows:
\begin{enumerate}
  \item \emph{State Prediction:}
  \begin{equation}\label{eq38}
    {{{\mathbf{\hat x}}}_{n\left| {n - 1} \right.}} = {\mathbf{g}}\left( {{{{\mathbf{\hat x}}}_{n - 1}}} \right),{{{\mathbf{\hat x}}}_{n + 1\left| {n - 1} \right.}} = {\mathbf{g}}\left( {{{{\mathbf{\hat x}}}_{n\left| {n - 1} \right.}}} \right).
  \end{equation}
  \item \emph{Linearization:}
  \begin{equation}\label{eq39}
    {{\mathbf{G}}_{n - 1}} = {\left. {\frac{{\partial {\mathbf{g}}}}{{\partial {\mathbf{x}}}}} \right|_{{\mathbf{x}} = {{{\mathbf{\hat x}}}_{n - 1}}}},{{\mathbf{H}}_n} = {\left. {\frac{{\partial {\mathbf{h}}}}{{\partial {\mathbf{x}}}}} \right|_{{\mathbf{x}} = {{{\mathbf{\hat x}}}_{n\left| {n - 1} \right.}}}}.
  \end{equation}
  \item \emph{MSE Matrix Prediction:}
  \begin{equation}\label{eq40}
    {{\mathbf{M}}_{n\left| {n - 1} \right.}} = {\mathbf{G}}_{n - 1}{{\mathbf{M}}_{n - 1}}{\mathbf{G}}_{n - 1}^H + {{\mathbf{Q}}_s}.
  \end{equation}
  \item \emph{Kalman Gain Calculation:}
  \begin{equation}\label{eq41}
    {{\mathbf{K}}_n} = {{\mathbf{M}}_{n\left| {n - 1} \right.}}{\mathbf{H}}_n^H{\left( {{{\mathbf{Q}}_m} + {{\mathbf{H}}_n}{{\mathbf{M}}_{n\left| {n - 1} \right.}}{\mathbf{H}}_n^H} \right)^{ - 1}}.
  \end{equation}
  \item \emph{State Tracking:}
  \begin{equation}\label{eq42}
    {{{\mathbf{\hat x}}}_n} = {{{\mathbf{\hat x}}}_{n\left| {n - 1} \right.}} + {{\mathbf{K}}_n}\left( {{{\mathbf{y}}_n} - {\mathbf{h}}\left( {{{{\mathbf{\hat x}}}_{n\left| {n - 1} \right.}}} \right)} \right).
  \end{equation}
  \item \emph{MSE Matrix Update:}
  \begin{equation}\label{eq43}
    {{\mathbf{M}}_n} = \left( {{\mathbf{I}} - {{\mathbf{K}}_n}{{\mathbf{H}}_n}} \right){{\mathbf{M}}_{n\left| {n - 1} \right.}}.
  \end{equation}
\end{enumerate}

\emph{Remark 2:} In the prediction step, the predicted angle ${\hat \theta}_{n\left| {n - 1} \right.}$ is used for transmit beamforming at the RSU at the $n$th epoch, ${\hat \theta}_{n+1\left| {n - 1} \right.}$ is sent to the vehicle for receive beamforming at the $\left(n+1\right)$th epoch. In the tracking step, based on the received target echo $\mathbf{y}_n$, the RSU refines the predicted state $\mathbf{\hat x}_{n\left| {n - 1} \right.}$ to obtain $\mathbf{\hat x}_n$, which will be used as the input of the predictor for the next iteration. By iteratively performing prediction and tracking, the RSU is able to simultaneously sense and communicate with the vehicle.

\section{Numerical Results}
In this section, we present the numerical results to validate the effectiveness of the proposed techniques for both angle tracking and downlink communication. Unless otherwise specified, both the RSU and the vehicle operate at $f_c = 30 \text{GHz}$, and we use $\Delta T = 0.02\text{s}$ as the block duration, $\sigma^2 = {\sigma_C^2} = 1$ as the noise variances for radar and communication, and ${\tilde \alpha} = 1$ as the reference communication channel coefficient. For the state evolution noises, we set $\sigma_\theta = 0.02^\circ$, $\sigma_d = 0.2\text{m}$, $\sigma_v = 0.5\text{m/s}$ and $\sigma_\beta = 0.1$, respectively. Note that here the variances for the state evolution are small since they stand for the approximation errors in the evolution models, which are irrelevant to the actual SNR. Moreover, the difference between two adjacent states is small given the short time duration $\Delta T$. As a consequence, the state variances should be set small enough. For the measurement noise variance, we set $a_1 = 1$, $a_2 = 6.7 \times 10^{-7}$ and $a_3 = 2 \times 10^4$. The matched-filtering gain is assumed to be $G = 10$. Without loss of generality, the initial state of the vehicle is set to $\theta_0 = 9.2^\circ$, $d_0 = 25\text{m}$, $v_0 = 18\text{m/s}$, ${\beta _0} = \frac{{\sqrt 2 }}{2} + \frac{{\sqrt 2 }}{2}j$ and ${\tilde \alpha} = 25$. Note that here we set ${\tilde \alpha} = d_0$ such that the modulus of the initial channel coefficient $\alpha_0$ is 1, which is the same as the reflection coefficient $\beta_0$ used in the DFRC scheme.
\begin{figure}[!t]
\centering
\subfloat[]{\includegraphics[width=0.85\columnwidth]{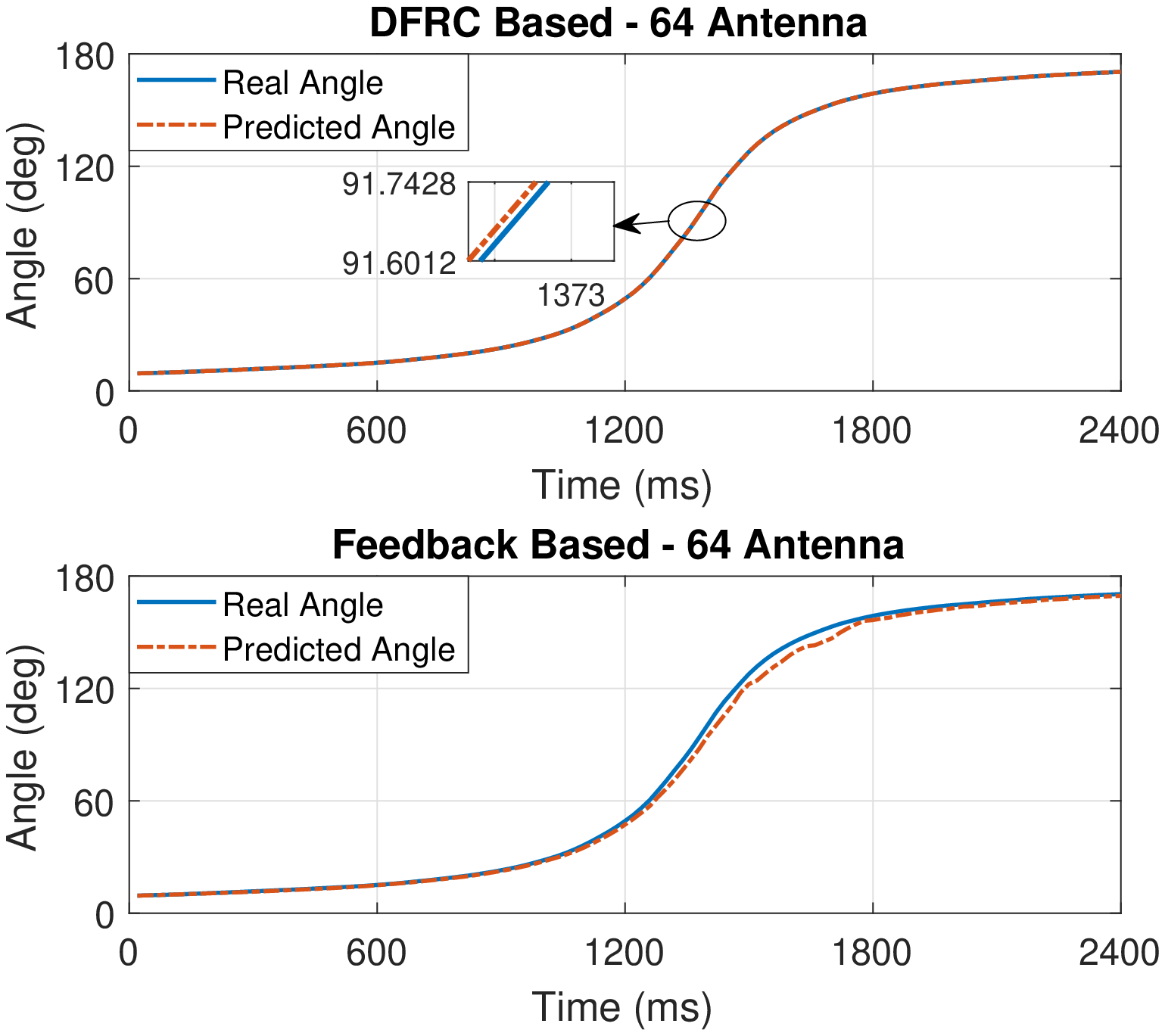}
\label{fig7a}}
\vspace{0.1in}
\hspace{.1in}
\subfloat[]{\includegraphics[width=0.85\columnwidth]{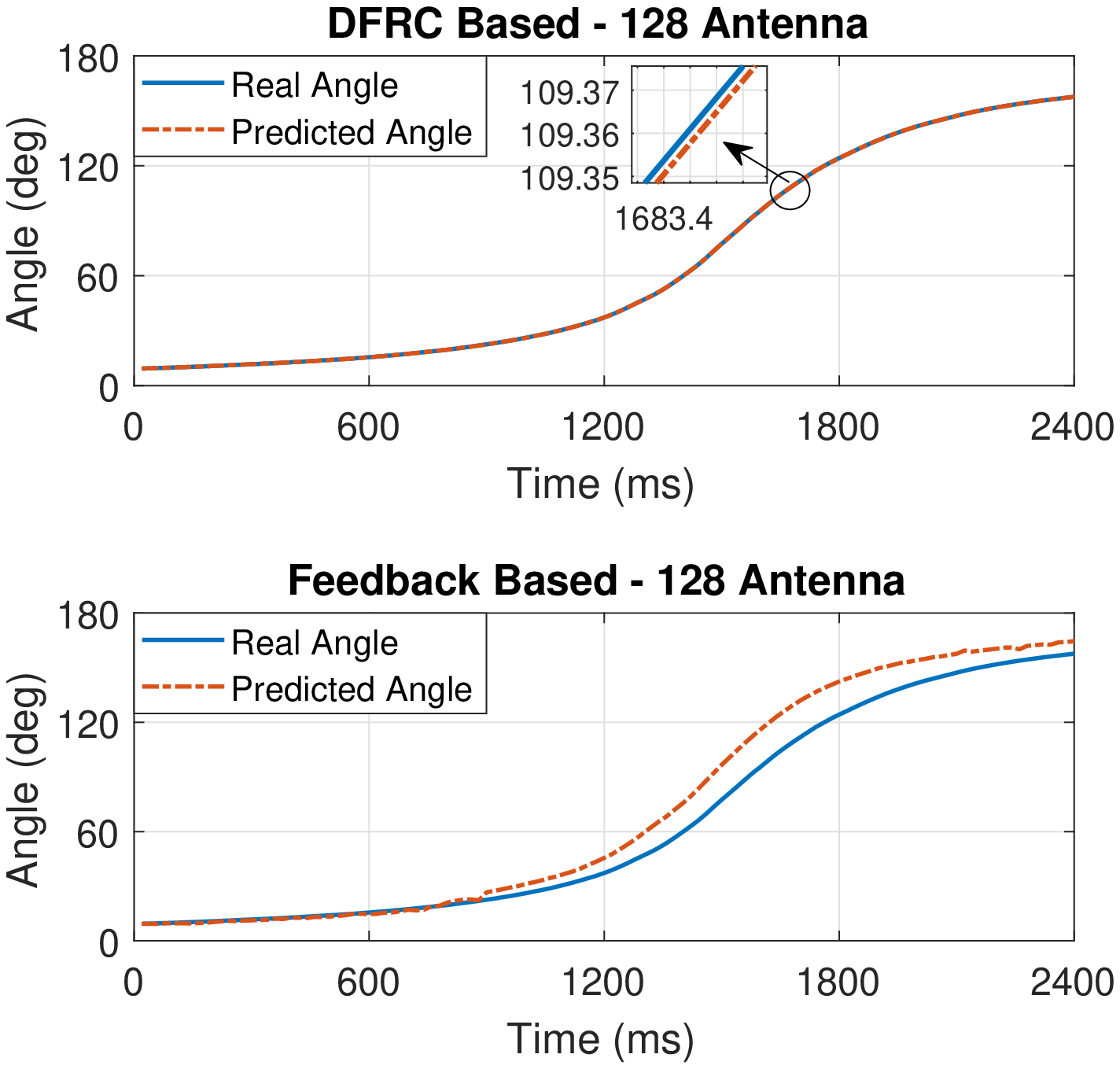}
\label{fig7b}}
\caption{Angle tracking performances for radar- and feedback-based schemes. (a) Initial state: $\theta_0 = 9.2^\circ$, $d_0 = 25\text{m}$, $v_0 = 18\text{m/s}$, ${\beta _0} = \frac{{\sqrt 2 }}{2} + \frac{{\sqrt 2 }}{2}j$, $\alpha_0 = 25$, $N_t = N_r = M = 64$, and transmit $\text{SNR} = 10\text{dB}$; (b) Initial state and SNR are the same as (a), antenna number $N_t = N_r = M = 128$.}
\label{fig7}
\end{figure}
\\\indent We compare the performance of the proposed DFRC-based beam tracking scheme and the benchmark communication-only feedback-based method. In conventional EKF based beam tracking schemes such as \cite{7905941}, the transmitter sends a single pilot vector to the receiver at each epoch. The receiver then combines the pilot by a receive beamformer, estimates the angle and feeds it back to the transmitter, which is used for predicting the transmit beam of the next time-slot. Note that all the existing EKF based beam tracking schemes employ state evolution models that are different to that of our paper \cite{7905941}. For the sake of fairness, we use the same state evolution model in the feedback based scheme for comparison, except that the reflection coefficient $\beta$ is now replaced by the LoS channel coefficient $\alpha$, which is assumed to be perfectly known. The angle measurement model for the feedback scheme is based on (\ref{eq11}), with the pilot symbol being matched-filtered. The distance and the velocity measurements are based on the time delay and the Doppler shift as well. Since there is only one single pilot being employed for tracking in the feedback based method, the matched-filtering gain is $G = 1$ in contrast to the DFRC case where $G = 10$. As a consequence, the measurement variances for the feedback-based approach are at an order of magnitude that is 10 times of that of the proposed DFRC scheme. Finally, we assume $N_t = N_r = M$ for fair comparisons.
\begin{figure}[!t]
    \centering
    \includegraphics[width=0.85\columnwidth]{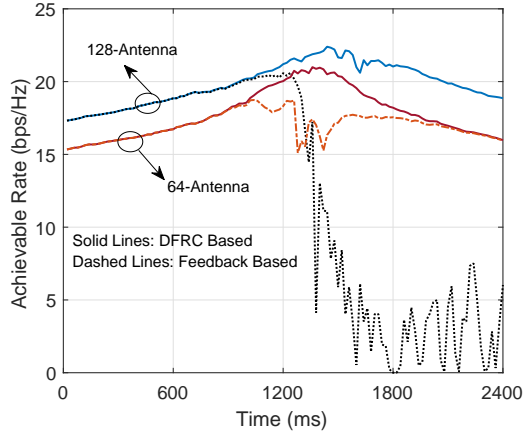}
    \caption{Achievable rate performances for radar- and feedback-based schemes, with initial state $\theta_0 = 9.2^\circ$, $d_0 = 25\text{m}$, $v_0 = 18\text{m/s}$, ${\beta _0} = \frac{{\sqrt 2 }}{2} + \frac{{\sqrt 2 }}{2}j$ and ${\tilde \alpha} = 25$, and transmit $\text{SNR} = 10\text{dB}$.  }
    \label{fig:8}
\end{figure}
\\\indent We first look at the angle tracking performance in Fig. 2(a) and (b). In the considered scenario, the vehicle starts from one side of the RSU, then passes in front of the RSU to its other side. Thanks to the matched-filtering gain in the DFRC based technique, both of the figures reveal that the proposed scheme can accurately track the variation of the vehicle's angle, while the feedback-based scheme shows larger tracking errors. Another important reason for this is that the feedback based approach requires the vehicle to use a receive beamformer to combine the pilot signal, which projects the pilot signal to a lower-dimensional space. This inevitably causes the loss of the angular information. On the other hand, the proposed DFRC scheme does not perform receive beamforming for the reflected echoes, which preserves most of the angular information. It is interesting to see from Fig. 2(b) that when the antenna array has larger size, the tracking error for the feedback based approach goes up, since the beam becomes narrower and the added SNR gain is not sufficient.
\\\indent In Fig. 3, we show the achievable rates for both DFRC and feedback based techniques. It can be observed that at the beginning when the angle variation is relatively slow, the feedback technique leads to almost the same rate performance as that of the proposed method. When the vehicle is approaching the RSU, however, the angle begins to vary rapidly, and the rate of the feedback method decreases drastically, which is consistent with the associated angle tracking performance in Fig. 2. Moreover, it can be observed in the 64-antenna case that the rate of the feedback based method catches up with that of the DFRC method when the vehicle is driving away. This is because when the angular variation is slow, the angle tracking error becomes acceptable, and the EKF for the feedback-based approach is still able to correct the angle deviation as shown in Fig. 2(a). Nevertheless, it fails to do so in the 128-antenna scenario given the narrower beam and higher misalignment probability, as depicted in Fig. 2(b).

\section{Conclusion}
In this paper, we have developed a novel predictive beamforming design for the V2I link by leveraging the joint sensing and communication capability deployed on the road side unit. Aiming for tracking the angular variation of the vehicle, we have proposed an extended Kalman filtering framework that builds upon the observation of the echo signal as well as the state evolution model of the vehicle. Numerical results have been provided to validate the proposed techniques, which have shown that the dual-functional radar-communication based beamforming design significantly outperforms the communication-only feedback-based schemes.





%
\bibliographystyle{IEEEtran}
\bibliography{IEEEabrv,Veh_REF}

\end{document}